\documentclass{PoS}

\title{Tevatron top physics}

\ShortTitle{Tevatron top physics}

\author{\speaker{Jeannine WAGNER-KUHR}\\
on behalf of the CDF and D\O\ Collaborations\\
        University of Karlsruhe, Karlsruhe, Germany\\
        E-mail: \email{jwagner@ekp.uni-karlsruhe.de}}

\abstract{
A summary of the most recent results on top 
quark physics obtained at Fermilab's Tevatron 
proton-antiproton collider, operating at a 
centre of mass energy of 1.96 TeV, is presented.
Measurements of the top pair and single top quark 
production cross sections, the investigation 
of top quark decay properties, the precision
measurement of the top quark mass as well
as searches for physics beyond the standard 
model are discussed.
}

\FullConference{Physics at LHC 2008\\
		 29 September - October 4, 2008\\
		 Split, Croatia}

\begin{document}

\section{Introduction}
In 1995 the top quark was discovered at the Tevatron 
proton-antiproton collider at Fermilab by the CDF
and D\O\ collaborations~\cite{topdis}. 
It is the most massive known elementary particle and
its mass of the order of the electroweak symmetry breaking
scale suggest that it may play a special role in
new physics models. So far the Tevatron collider is the 
only place to study the top quark.\\
At the Tevatron collider, with a center of mass energy 
of $\sqrt{s}=1.96\;\mbox{TeV}$, most top quarks are 
pair produced via the strong interaction. About 
85\% of the top pairs are produced via 
quark-antiquark annihilation and 15\% via gluon fusion.
Beside the production of top quarks in pairs also
the production of single top quarks in electroweak
interaction should exist and its cross section is 
predicted to be $\sim 2.5$ smaller than the top pair 
production cross section.\\
In the standard model the top quark decays predominantly 
into a $W$ boson and a $b$ quark, with a branching ratio
close to 100\%. Different decay channels are distinguished 
according to the decay mode of the $W$ boson.\\
Top pair events with both $W$-bosons decaying leptonically to
$e\nu$ or $\mu\nu$ are called dilepton events, events with
only one $W$-boson decaying leptonically are called 
lepton+jets events and events with both $W$-bosons decaying
hadronically are called all hadronic events.
Most of the top pair measurements are carried
out in the lepton+jets channel, which is characterized by moderate
background and a branching ratio of about 30\%. In
single top quark events we consider only the leptonic
decay of the $W$-boson, the branching ratio is here about 21\%.

\section{Top Quark Properties in Production and Decay}
Investigating top quark properties in production
and decay will answer the question whether the observed 
top quark behaves actually like the top quark predicted 
in the standard model (SM).
In this article the production cross section 
for top pairs and single top quarks, 
characteristics of top pair production like the 
fraction of gluon fusion production and
the forward-backward charge asymmetry, and 
the Wtb coupling are discussed.

\subsection{Single Top Quark Production Cross Section}
\begin{figure}
\includegraphics[width=0.32\textwidth]{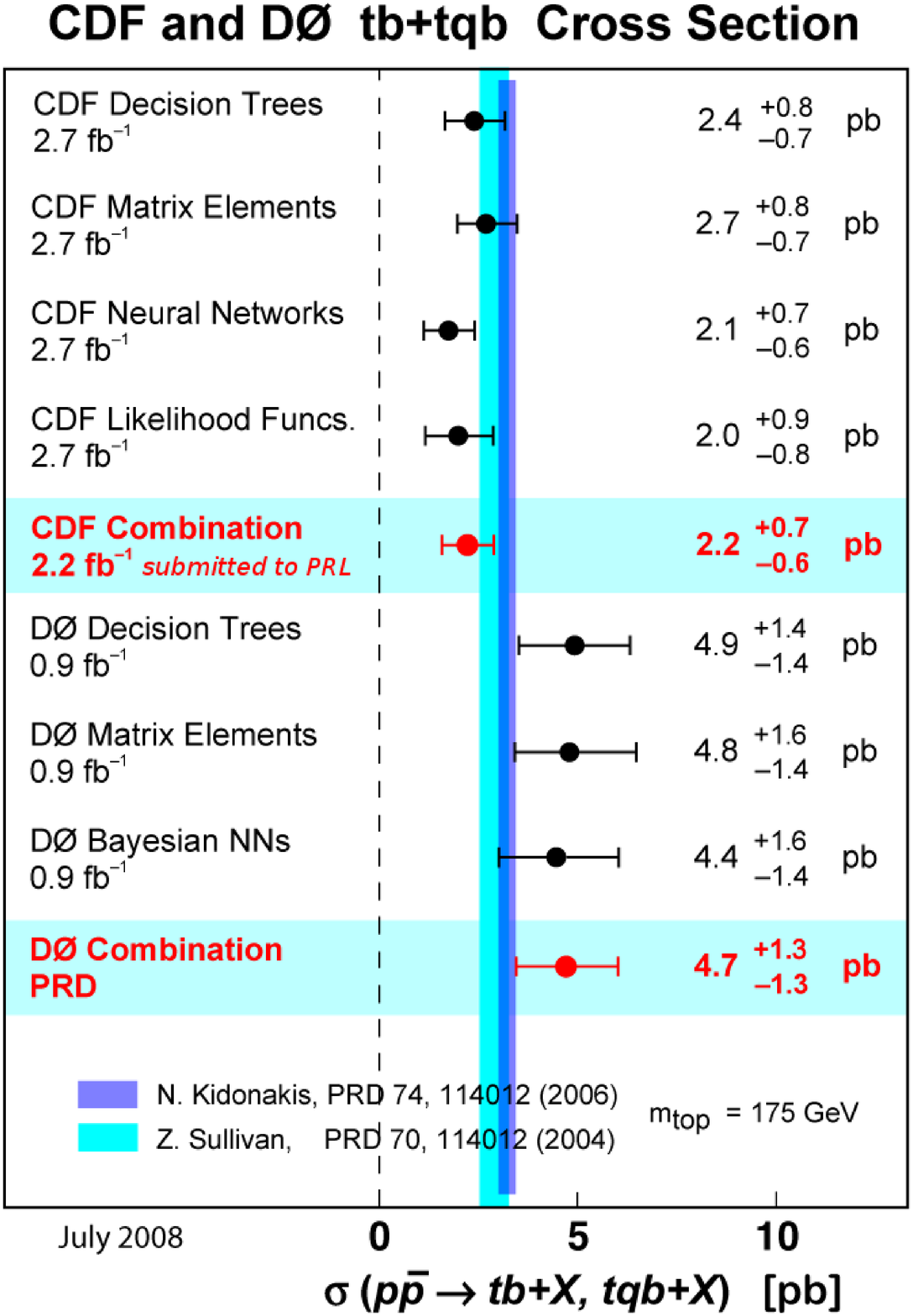}
\includegraphics[width=0.32\textwidth]{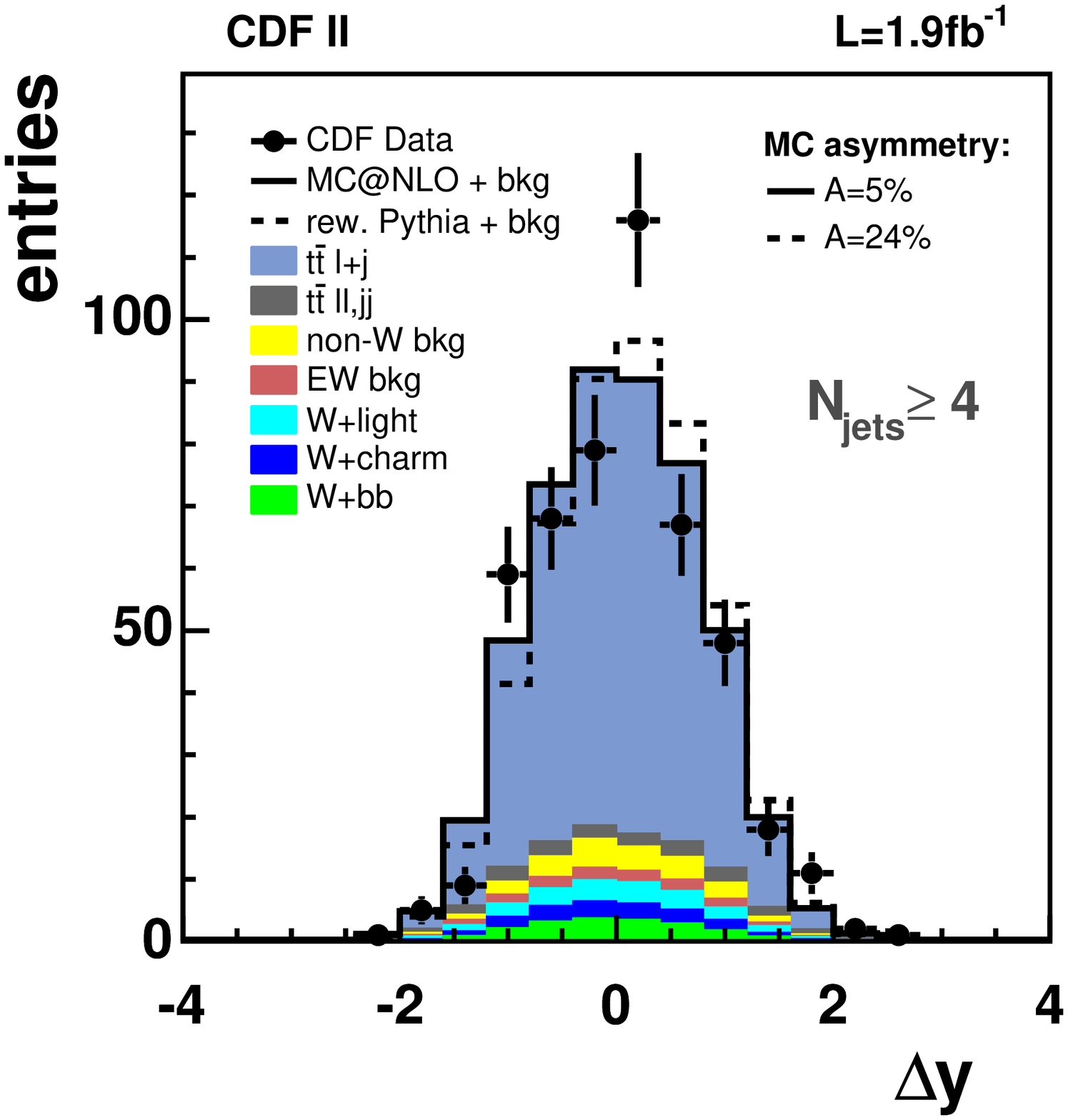}
\includegraphics[width=0.34\textwidth]{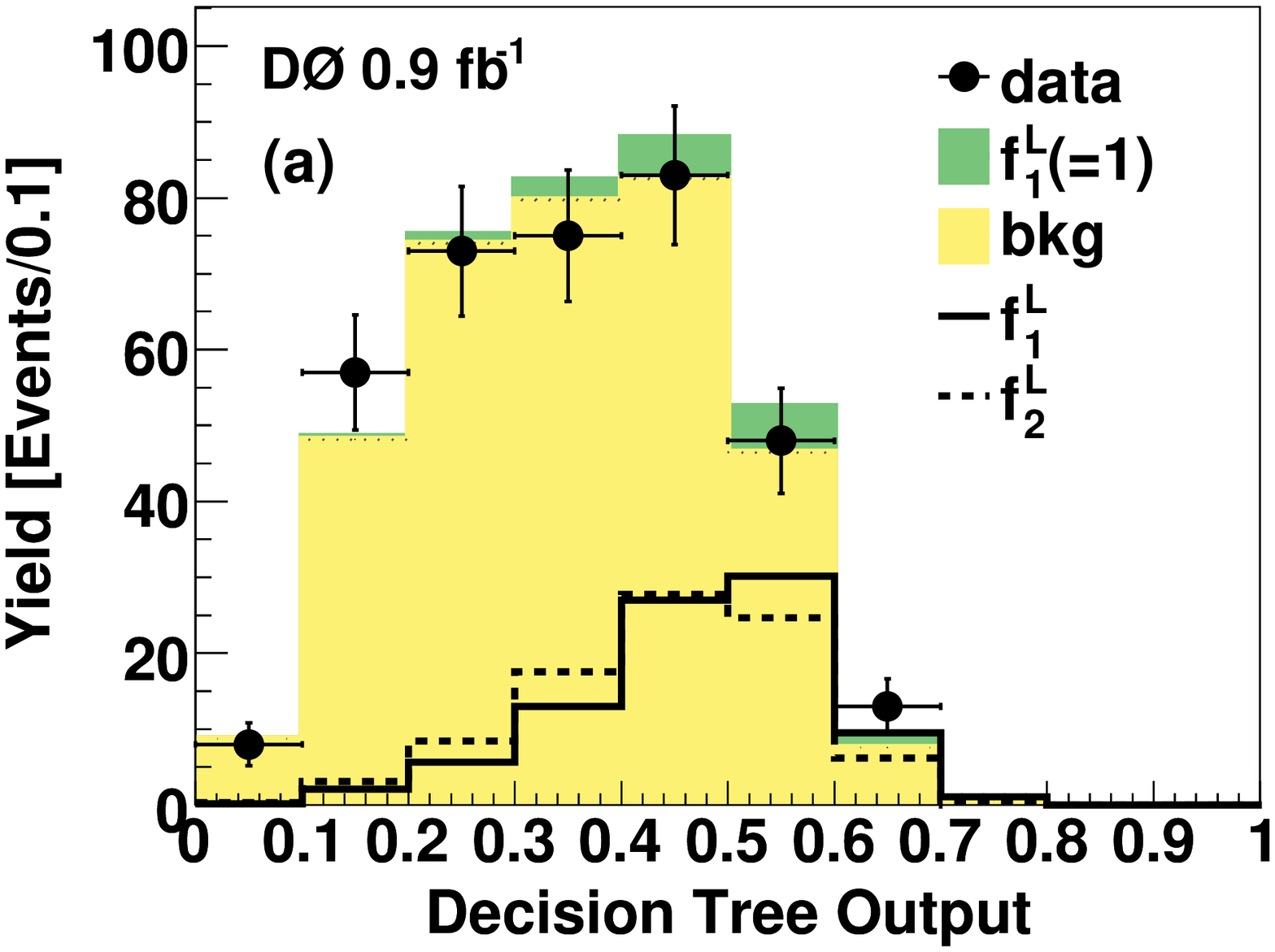}
\caption{Left: Single top quark production. Shown is an overview of the cross section measurements performed at the Tevatron.
Middle: Measurement of the forward-backward charge asymmetry. 
Shown is the reconstructed rapidity difference of the top 
and anti-top quark. Right: Study of the $Wtb$ coupling. 
Shown is the single top quark boosted decision tree
discriminant assuming $f_1^R=f_2^R=0$.}
\label{fig1}
\end{figure}

The challenge of measuring the single top quark cross
section at the Tevatron is the huge background of
W+jets events. A simple cut based 
event selection is here not sufficient and to improve
the sensitivity both experiments use multivariate 
techniques, like likelihood functions, decision trees, 
matrix element methods and neural networks. An overview
of the most recent single top quark cross section measurements
performed at the Tevatron is presented in figure~\ref{fig1}.
The CDF likelihood function, neural network and matrix element discriminants 
have been combined in a neural network to obtain a super discriminant.
An analysis employing this technique on $2.2\,\mbox{fb}^{-1}$ 
saw an excess of signal over background, and a cross section of
$2.2^{+0.7}_{-0.6}\;\mbox{pb}$ is obtained \cite{ST_CDF}. 
The result is below but consistent with
the SM prediction of $2.9\pm 0.4\;\mbox{pb}$. 
The expected signal significance of the CDF
combination is $5.1\,\sigma$ while a significance of $3.7\,\sigma$ is observed.
The three multivariate analyses~\cite{ST_DO} from D\O\ have been combined 
using BLUE, the best linear unbiased estimator. A cross section of 
$4.7\pm 1.3\,\mbox{pb}$
is obtained, which is larger than but consistent with 
the SM prediction. The expected signal significance of the D\O\ combination is 
$2.3\,\sigma$ while a significance of $3.6\,\sigma$ has been observed.
Both Tevatron experiments see strong evidence for electroweak 
single top quark production.

\subsection{Top Pair Production}
The top pair production cross section has been measured in different
decay channels and with different methods. 
In contrast to the single top quark production measurement, many 
top pair cross section measurements are counting experiments, as long
as they are performed in regions with low background (e.g. lepton+jets
channel with 4 jets and at least one jet tagged as $b$-jet).
The most precise measurements are obtained in the lepton+jets channel.
All top pair cross section measurements are consistent with each other 
and assuming a top quark mass of $175\,\mbox{GeV/c}^2$ the combination 
of the CDF measurements yields a cross
section of $7.0\pm 0.3\,\mbox{(stat.)}\pm 0.4\,\mbox{(syst.)}\pm 0.4\,\mbox{(lumi)}\,\mbox{pb}$, while the combination of the D\O\ results yields
$7.8\pm 0.5\,\mbox{(stat.)}\pm 0.6\,\mbox{(syst.)}\pm 0.5\,\mbox{(lumi)}\,\mbox{pb}$ \cite{TT_sigma}.
Thus, the top pair production cross section is now known
with a relative uncertainty of about 9\% at Tevatron center of
mass energies.

The fraction of top pair events produced via gluon fusion has
been measured in three CDF analyses~\cite{TT_fgg}. 
Two analyses use variables which are sensitive to the different
spin configuration of the top quarks in gluon fusion and 
quark-antiquark annihilation events. The analysis performed
in the lepton+jets uses a neural network while the more
recent analysis in the dilepton channel utilizes just the azimuthal
angle difference of the two charged leptons.
A third analysis, performed in the lepton+jets channel,
exploits the fact that gluons emit more soft gluons than
quarks and so the number of tracks with small transverse momentum
is used as sensitive variable. The combination of both lepton+jets 
analyses yields a gluon fusion fraction of $(7^{+15}_{-7})\%$ ($1\,\mbox{fb}^{-1}$), 
while the dilepton analysis obtains a fraction of $(53^{+36}_{-38})\%$
($2\,\mbox{fb}^{-1}$).
Both results are consistent with the SM prediction of about 15\%.

At the Tevatron a small forward-backward asymmetry of about 5\%
is predicted in SM top pair production. In new physics scenarios
with a $Z^\prime$ or an axigluon this asymmetry can be as large as
$\pm 30\%$. The asymmetry is frame dependent and is predicted to
be largest in the top pair rest frame. To measure the asymmetry 
the lepton+jets channel is used and the asymmetry in the top pair 
rest frame has been studied by both experiments, using 
the rapidity difference of the top and anti-top. This quantity is 
presented in figure~\ref{fig1} for the CDF analysis. The CDF
collaboration performed also an asymmetry measurement in the
laboratory frame using the polar angle of the hadronically 
decaying top quark
as sensitive variable. The D\O\ collaboration obtains an 
uncorrected asymmetry of $(12\pm 8)\%$~\cite{Afb_DO}, 
while an asymmetry of $0.8\%$ is predicted. 
To account for sizable acceptance and measurement
dilution effects the CDF collaboration corrects the raw
results for these effects leading to an asymmetry of $(24\pm 14)\%$
in the top rest frame and to $(18\pm 8)\%$ in the laboratory frame~\cite{Afb_CDF}.
The measured asymmetries at the Tevatron are larger but consistent
with the SM prediction within uncertainties.

\subsection{$Wtb$ coupling}
The general $Wtb$ vertex contains besides the left-handed
vector coupling $f_1^L$ also a right-handed 
vector coupling $f_1^R$ and tensor couplings $f_2^L$ and $f_2^R$.
In the SM only $f_1^L$ is non-zero, namely one, and non-SM
couplings would affect the single top quark production in
rate and kinematic distributions. The D\O\ collaboration
uses the single top quark boosted decision tree discriminant
to determine two couplings at a time while the other two 
are assumed to be negligible, see figure~\ref{fig1}.
This analysis yields for the scenario with $f_1^R=f_2^R=0$
the couplings $f_1^L=1.4^{+0.6}_{-0.5}$ and 
$f_2^L<0.5$ at 95\% C.L. \cite{ST_Wtb}.\\
Besides single top quark production also the helicity of
$W$-bosons in top pair decays would be affected by non-SM 
$Wtb$ couplings.  In the SM the fraction of longitudinally 
polarized $W$ bosons is $f_0=0.7$, the fraction of left-handed 
$W$-bosons $f_-=0.3$ and the fraction of right-handed W-bosons 
is $f_+=0.0$. The reconstructed $\cos\theta^*$, with $\theta^*$
being the angle between the charged lepton or the down 
type quark and the negative direction of the top quark in the rest 
frame of the W boson, is here used as sensitive variable.  
The most recent CDF analyses use the leptonically decaying
$W$ boson in the lepton+jet channel and
two analyses have been combined with BLUE. 
The D\O\ analysis uses also the dilepton channel and uses both
$W$ bosons in each event in case of the lepton+jets channel.
The reconstructed $\cos\theta^*$ distribution is fitted and
the CDF combination yields for the simultaneous fit $0.66\pm 0.16$ 
for $f_0$ and $-0.03\pm 0.07$ for $f_+$ while the D\O\ collaboration
obtains $0.49\pm 0.13$ for $f_0$ and $0.11\pm 0.07$ for $f_+$
\cite{TT_Wtb}.
All $W$-helicity measurements in top pair events performed
at the Tevatron are compatible with the SM prediction.

\section{Top Mass Measurement}
\begin{figure}
\hspace{0.05\textwidth}
\includegraphics[width=0.30\textwidth]{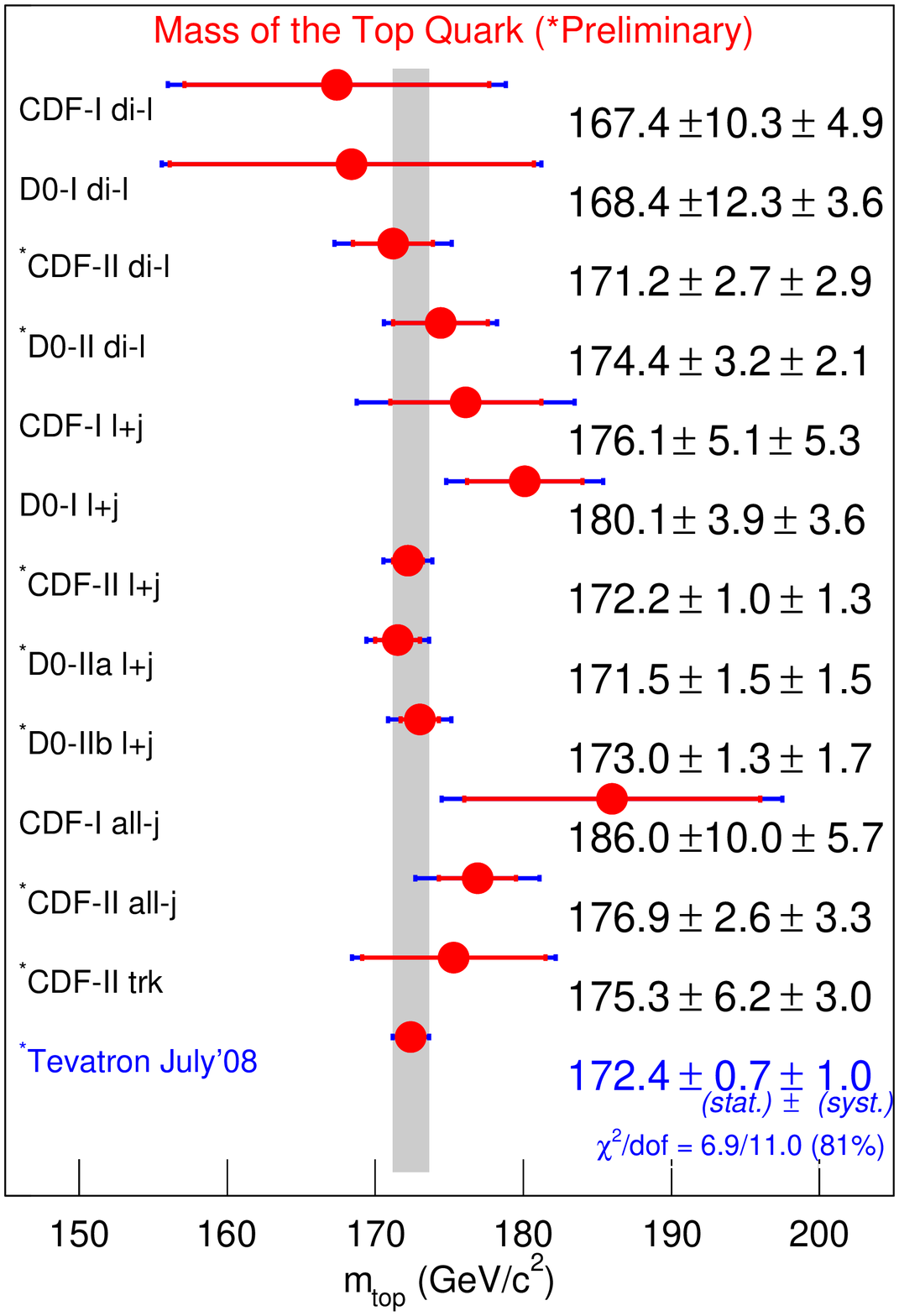}
\hspace{0.05\textwidth}
\includegraphics[width=0.6\textwidth]{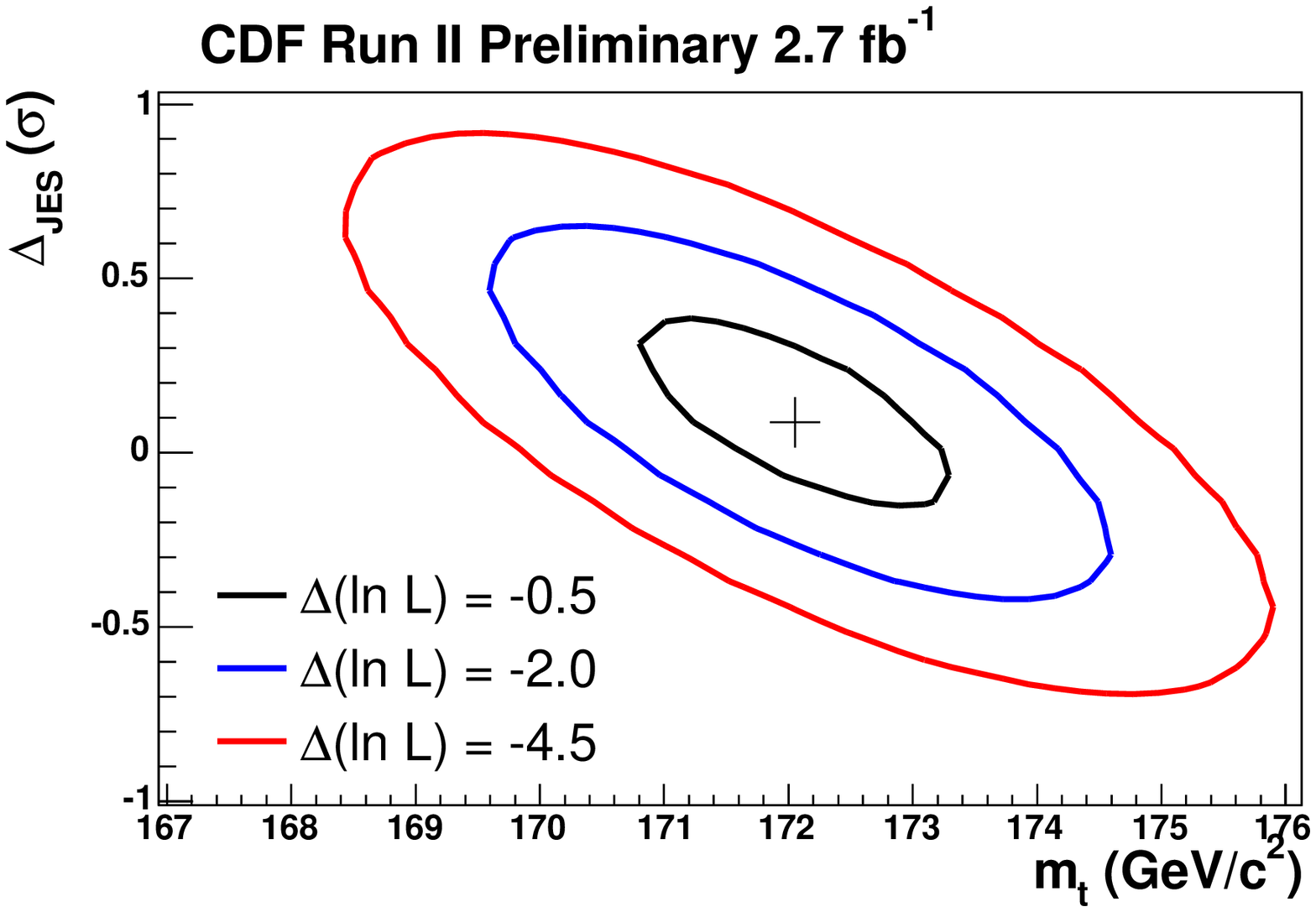}
\caption{Measurement of the top quark mass. Left: Tevatron top mass combination. Right: Most precise top mass measurement in the lepton+jets channel performed by the CDF collaboration.}
\label{fig2}
\end{figure}
The mass $m_t$ of the top quark is an important standard model 
parameter and precision measurements of the top quark and $W$ 
masses test the consistency of the SM, and in particular the 
Higgs mechanism.
In order to measure the top quark mass top pair events
in all decay channels are used and to reduce the systematic 
uncertainty due to the jet energy scale uncertainty the jet
energy scale (JES) is calibrated in an in situ measurement using 
the hadronically decaying $W$ boson in the lepton+jets and 
all hadronic channel. The CDF collaboration has also performed 
measurements in the lepton+jets channel using not the reconstructed 
top mass as sensitive variable but the transverse momentum of the 
charged lepton and the decay length of $b$-tagged jets. 
These variables are almost entirely independent of the 
jet energy scale and thus provide a cross check on the
measurements involving JES, albeit statistically limited. 
The combination of all these measurements, see figure~\ref{fig2},
yields a top mass of $172.4\pm 0.7\,\mbox{(stat.)}\pm 1.0\,\mbox{(syst.)}
\,\mbox{GeV/c}^2$~\cite{mt_comb},
meaning that the top quark mass is now known with a relative precision 
of 0.7\%. The measurements performed in all channels and with
different methods are consistent with each other.\\
The most precise top quark mass measurements are carried out
in the lepton+jets channel and the most recent lepton+jets 
measurements use the matrix element method, the in situ jet 
energy scale calibration and neural networks. Figure~\ref{fig2}
shows for the CDF measurement the likelihood contours as a 
function of the top mass and the in situ jet energy scale parameter.
The most precise single measurements of both experiments yield a 
top mass of $172.2\,\mbox{GeV/c}^2$ with a relative uncertainty 
of about 1\%~\cite{mt_prec}. 

\section{Search for New Physics in the Top Quark Sector}
Since the top quark is the only known fermion 
with a mass of the order of the electroweak symmetry breaking 
scale it may play a special role in new physics models.
Searches for new physics in top quark events and in
events with top quark signature (e.g. lepton+jets channel)
are discussed.

Both Tevatron experiments performed searches for heavy
massive resonances in top pair~\cite{TT_resonances} and 
single top quark production~\cite{ST_resonances}. 
In case of top pair events the reconstructed
mass of the top quark pair system is studied and D\O\ 
and CDF find so far no hint for a heavy resonance, e.g. $Z'$, 
heavy gluon. The D\O\ collaboration sets for example a lower
limit on a narrow leptophobic $Z^\prime$ of $760\,\mbox{GeV/c}^2$.
To search for resonant single top quark production (e.g. 
charged Higgs, $W^\prime$) the invariant mass of the reconstructed 
top quark and the additional $b$-jet is used. Both Tevatron
experiments find so far no hints for massive resonances
in single top quark production.

\begin{figure}
\hspace{0.05\textwidth}
\includegraphics[width=0.4\textwidth]{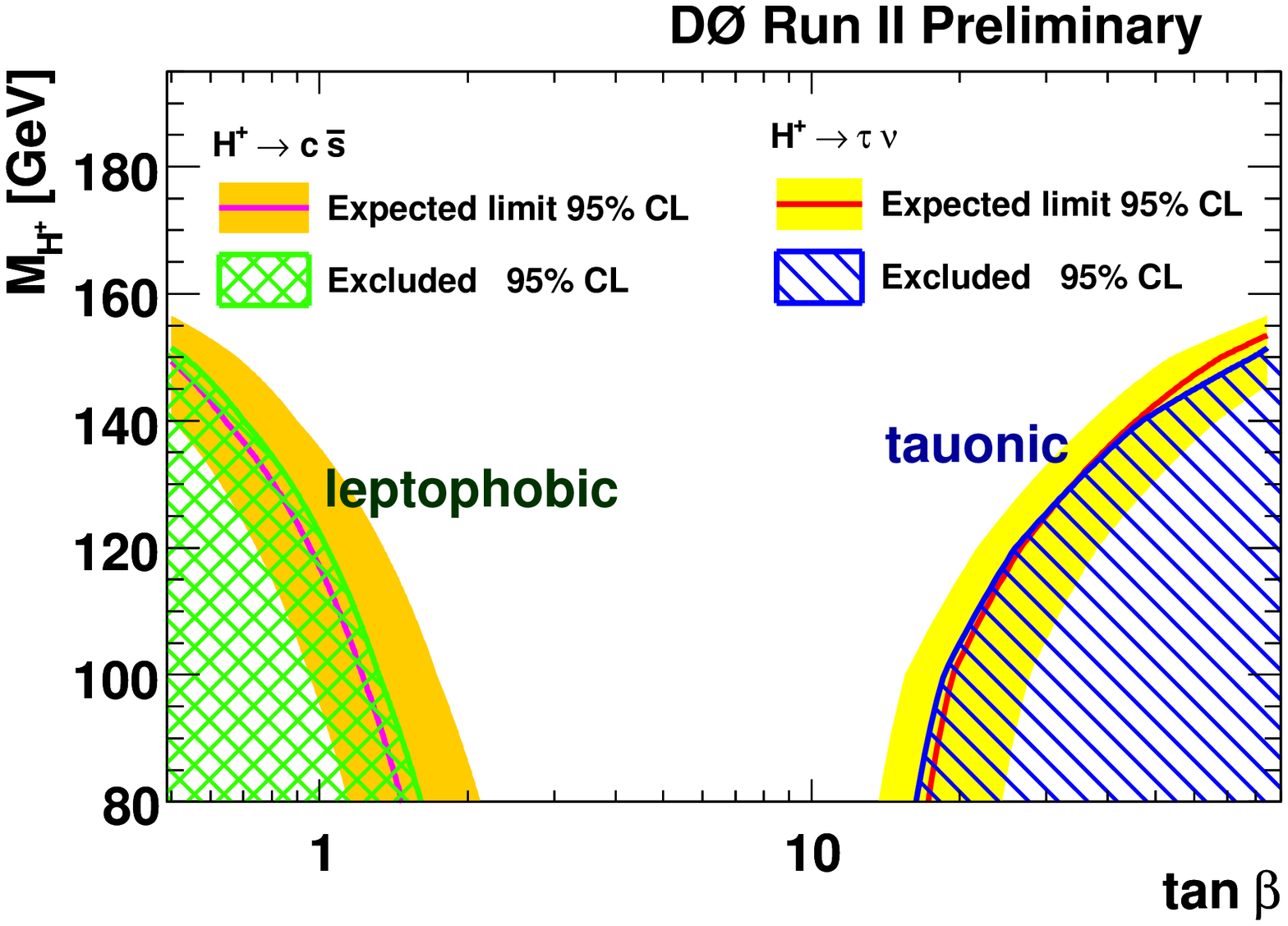}
\hspace{0.08\textwidth}
\includegraphics[width=0.38\textwidth]{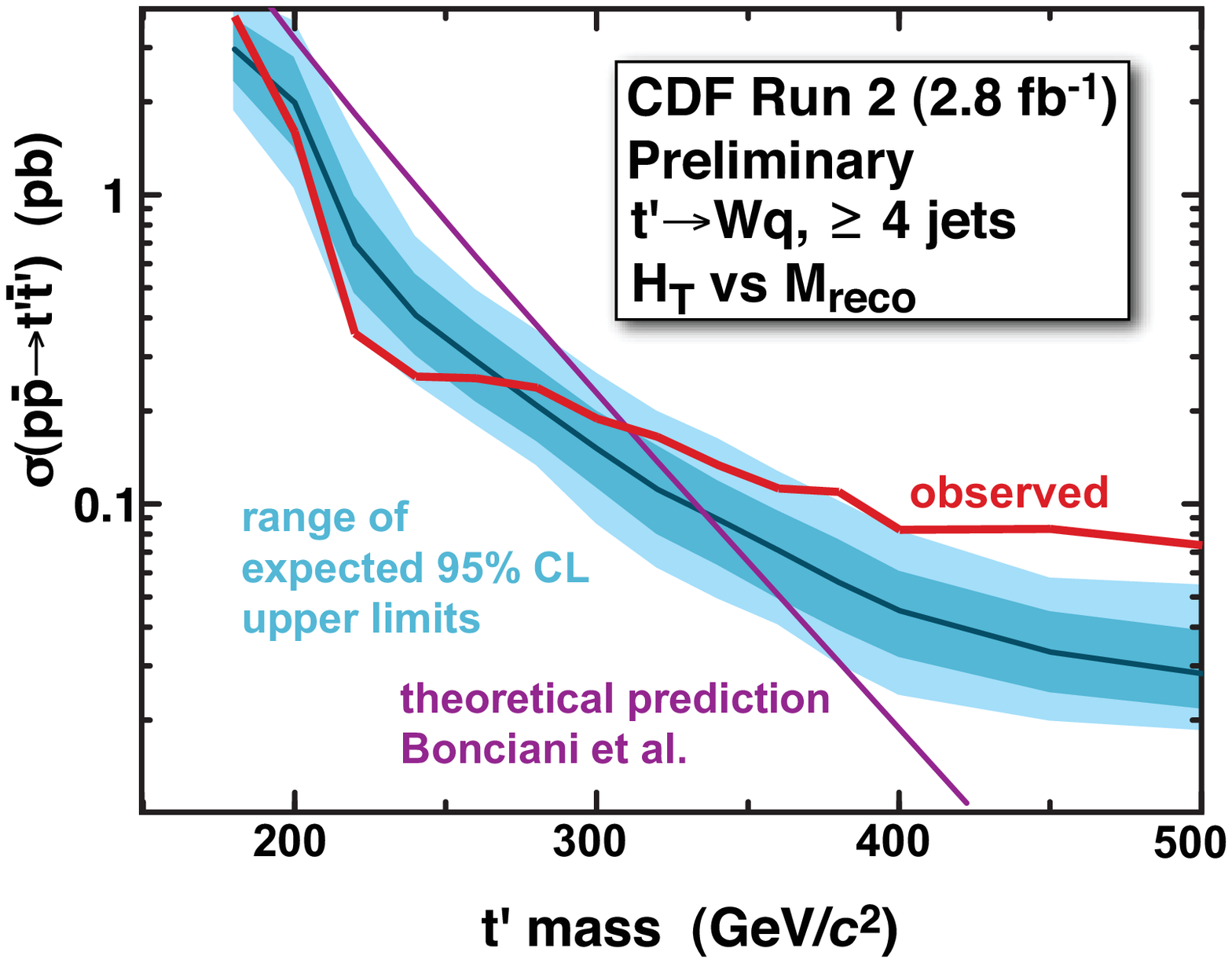}
\caption{Search for new physics in the top quark sector. Left: Search for a charged Higgs in the decay of top pair events. Shown is the limit as a function
of the charged Higgs mass and $\tan\beta$ obtained by the D\O\ analysis.
 Right: Search for $t^\prime$ pair production. Shown is the limit on the $t^\prime$ production cross section as a function of the reconstructed $t^\prime$ mass.}
\label{fig3}
\end{figure}
The CDF and D\O\ collaboration have also performed searches for a 
charged Higgs in the decay of top pair events \cite{tt_Higgs}. The decay 
$t\rightarrow bH^+$ is allowed in the minimal super-symmetric SM
(MSSM). At low $\tan\beta$ the decay $H^+\rightarrow c\bar{s}$
is important while at large $\tan\beta$ the decay $H^+\rightarrow\tau\nu$
becomes dominant. Here $\tan\beta$ is the ratio of the
vacuum expectation values of the two Higgs doublets.
The CDF collaboration searches for $H^+\rightarrow c\bar{s}$
by using the invariant mass of the two light-quark jets assigned
as jets originating from the charged Higgs.
The D\O\ collaboration chose a different ansatz and searches
for a charged Higgs decaying into either a $c\bar{s}$ or $\tau\nu$
by measuring the rate of top pair events across the decay channels: 
lepton+jets, dilepton and tau+lepton. The limits as a function
of the charged Higgs mass and $\tan\beta$ obtained by the 
D\O\ analysis are presented in figure~\ref{fig3}.

Most recent searches for flavor changing neutral currents (FCNC)
have been performed by the CDF collaboration. In top pair
events CDF searches for FCNC decays like $t\rightarrow Zq$
by looking at the $\chi^2$ of a kinematic fit with FCNC hypothesis.
From a fit to this $\chi^2$ variable an upper limit of 3.7\% for the 
branching ratio $t\rightarrow Zc$ is derived \cite{TT_FCNC}. 
The CDF collaboration
searches also for FCNC in single top quark production and looks
here for events where the fusion of a gluon and a $u$- or $c$-quark
leads to a single top quark. A neural network is used to separate 
FCNC top quark events from background events. From a fit to
the neural network output an upper limit on the anomalous
single top quark cross section of $1.8\,\mbox{pb}$ is determined
and upper limits on the anomalous couplings of 
$\kappa_{gtu}=0.018\,\mbox{TeV}^{-1}$ and 
$\kappa_{gtc}=0.069\,\mbox{TeV}^{-1}$ are derived \cite{ST_FCNC}.

Searches for new physics in events with top quark signature
have been performed by both Tevatron experiments. Here, only
the most recent results are discussed. The CDF collaboration
searches in events with top pair lepton+jets signature
for a heavy top like quark~\cite{tprime}, a $t^\prime$, decaying into 
a $W$ boson and a quark ($b,s,d$) assuming that the mass 
split between a $b^\prime$ and a $t^\prime$ is below the $W$ boson mass. 
The reconstructed $t^\prime$ mass as well as the total transverse 
energy in the event are used in a 2-dimensional likelihood fit
leading to the limit on the $t^\prime$ production cross section 
presented in figure~\ref{fig3}. In events with top pair dilepton
signature CDF searches for a light super-symmetric top partner ($\lesssim m_t$), 
the stop~\cite{stop}. In this analysis, it is assumed that the stop decays 
with 100\% into a b quark and a chargino which would then decay 
via different modes into a neutralino, a charged lepton, and a neutrino.
The stop mass is reconstructed for each event assuming
the pair of neutralino and neutrino to be a pseudo particle
and limits are extracted from this variable showing that
there is so far no hint for a stop in the data.

\section{Conclusion}
In this article the most recent results on top quark physics from
the Tevatron have been presented. Many efforts have been
made to study the top quark and so far the top quark behaves 
as predicted in the standard model. The top quark mass is measured
very precisely with a relative precision of 0.7\% and so far
no significant hints for new physics in top quark events as well 
as in events with top quark signature have been found.
Most measurements of top quark properties are limited by statistics,
so we expect these to continue to improve in precision along
with the sensitivity to new physics as more data is collected.


\begin{thebibliography}{99}
\bibitem{topdis} F. Abe {\it et al.} (CDF Collaboration), Phys.~Rev.~Lett. {\bf 74}, 2626 (1995); S. Abachi {\it et al.} (D\O\ Collaboration), Phys.~Rev.~Lett. {\bf 74}, 2632 (1995).
\bibitem{ST_CDF} T. Aaltonen {\it et al.} (CDF Collaboration), Phys.~Rev.~Lett. {\bf 101}, 252001 (2008); public CDF notes 9479, 9451, 9464, 9445.
\bibitem{ST_DO} V.M. Abazov {\it et al.} (D\O\ Collaboration), Phys.~Rev. D{\bf 78}, 012005 (2008).
\bibitem{TT_sigma} Public CDF note 9448; public D\O\ note 5715-CONF.
\bibitem{TT_fgg} T. Aaltonen {\it et al.} (CDF Collaboration), submitted to Phys.~Rev.~Lett., arXiv:0807.4262; T. Aaltonen {\it et al.} (CDF Collaboration), Phys.~Rev. D{\bf 78},111101 (2008); public CDF note 9432;
\bibitem{Afb_DO} V.M. Abazov {\it et al.} (D\O\ Collaboration), Phys.~Rev.~Lett. {\bf 100}, 142002 (2008).
\bibitem{Afb_CDF} T. Aaltonen {\it et al.} (CDF Collaboration), Phys.~Rev.~Lett. {\bf 101}, 202001 (2008).
\bibitem{ST_Wtb} V.M. Abazov {\it et al.} (D\O\ Collaboration), Phys.~Rev.~Lett. {\bf 101}, 221801 (2008).
\bibitem{TT_Wtb} Public D\O\ note 5722-CONF; T. Aaltonen {\it et al.} (CDF Collaboration), submitted to Phys. Lett. B, arXiv:0811.0344. 
\bibitem{mt_comb} Tevatron Electroweak Working Group and CDF and D\O\ Collaborations, arXiv:0808.1089.
\bibitem{mt_prec} Public CDF note 9427; public D\O\ note 5750-CONF.
\bibitem{TT_resonances} Public CDF note 9157; public CDF note 9164; public D\O\ note 5600-CONF. 
\bibitem{ST_resonances} V.M. Abazov {\it et al.} (D\O\ Collaboration), submitted to Phys.~Rev.~Lett., arXiv:0807.0859; V.M. Abazov {\it et al.} (D\O\ Collaboration), Phys.~Rev.~Lett. {\bf 100}, 211803 (2008); public CDF note 9150.
\bibitem{tt_Higgs} Public CDF note 9322; public D\O\ note 5715-CONF
\bibitem{TT_FCNC} T. Aaltonen {\it et al.} (CDF Collaboration), Phys.~Rev.~Lett. {\bf 101}, 192002 (2008).
\bibitem{ST_FCNC} T. Aaltonen {\it et al.} (CDF Collaboration), submitted to Phys.~Rev.~Lett., arXiv:0812.3400.
\bibitem{tprime} Public CDF note 9446. 
\bibitem{stop} Public CDF note 9439.
\end{thebibliography}
\end{document}